\documentclass[conference]{IEEEtran}

\IEEEoverridecommandlockouts
\usepackage{cite}
\usepackage{amsmath,amssymb,amsfonts}
\usepackage{algorithm}
\usepackage{algpseudocode}
\usepackage{graphicx}
\usepackage{textcomp}
\usepackage{amsmath}  
\usepackage{amsfonts}  
\usepackage{amssymb}  
\usepackage{soul}
\usepackage{enumitem}
\usepackage{xcolor}
\usepackage{booktabs}
\usepackage{cuted}
\usepackage{svg}
\usepackage{tcolorbox}
\usepackage{bm}
\usepackage{amsmath,amssymb}
\usepackage{algorithm}
\usepackage[letterpaper, right=0.622in, left=0.622in, top=0.73in, bottom=1.02in]{geometry}
\usepackage{subcaption} 

\def\BibTeX{{\rm B\kern-.05em{\sc i\kern-.025em b}\kern-.08em
    T\kern-.1667em\lower.7ex\hbox{E}\kern-.125emX}}

\makeatletter
\def\@fnsymbol#1{\ensuremath{\ifcase#1\or *\or \dagger\or \ddagger\or \mathsection\or \mathparagraph\or \|\or **\or \dagger\dagger\or \ddagger\ddagger \fi}}
\makeatother

\begin{document}

\title{MA-Aided Hierarchical Hybrid Beamforming for Multi-User Wideband Beam Squint Mitigation}
\author{
\IEEEauthorblockN{Cixiao Zhang, Yin Xu, Xinghao Guo, Xiaowu Ou, Dazhi He and Wenjun Zhang}

\IEEEauthorblockA{Cooperative Medianet Innovation Center (CMIC), Shanghai Jiao Tong University, Shanghai 200240, China \\
Email: \{cixiaozhang, xuyin, guoxinghao, xiaowu\_ou, hedazhi, zhangwenjun\}@sjtu.edu.cn}


\thanks{
The corresponding author is Yin Xu (e-mail: xuyin@sjtu.edu.cn).}
}

\maketitle

\begin{abstract}
In wideband near-field arrays, frequency-dependent array responses cause wavefronts at different frequencies to deviate from that at the center frequency, producing beam squint and degrading multi-user performance. True-time-delay (TTD) circuits can realign the frequency dependence but require large delay ranges and intricate calibration, limiting scalability. Another line of work explores one- and two-dimensional array geometries, including linear, circular, and concentric circular, that exhibit distinct broadband behaviors such as different beam-squint sensitivities and focusing characteristics. These observations motivate adapting the array layout to enable wideband-friendly focusing and enhance multi-user performance without TTD networks. We propose a movable antenna (MA) aided architecture based on hierarchical sub-connected hybrid beamforming (HSC-HBF) in which antennas are grouped into tiles and only the tile centers are repositioned, providing slow geometric degrees of freedom that emulate TTD-like broadband focusing while keeping hardware and optimization complexity low. We show that the steering vector is inherently frequency dependent and that reconfiguring tile locations improves broadband focusing. Simulations across wideband near-field scenarios demonstrate robust squint suppression and consistent gains over fixed-layout arrays, achieving up to 5\% higher sum rate, with the maximum improvement exceeding 140\%.
\end{abstract}


\begin{figure*}[!t]
  \centering
  \includegraphics[width=\textwidth]{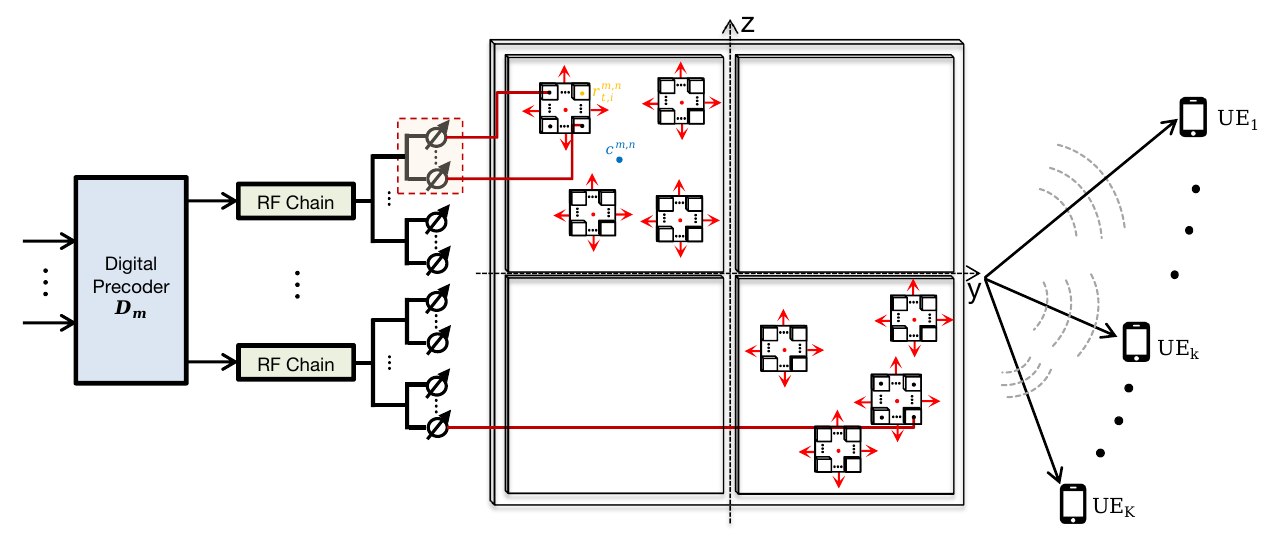}
  \caption{MA-aided hierarchical sub-connected hybrid beamforming structure.}
  \vspace{-0.5cm}
  \label{fig:system-model}
\end{figure*}

\section{Introduction}
To meet the stringent demands of high data rates, ultra-low latency, and ultra-reliability, beyond the fifth generation (5G) and sixth generation (6G) systems are envisioned to operate at millimeter-wave (mmWave) and terahertz (THz) frequencies with multi-gigahertz bandwidths and electrically large apertures \cite{6Gsurvey1,Thzsurvey1}. However, these high-frequency bands suffer from severe path loss that increases with carrier frequency. Extremely large-scale multiple-input multiple-output (XL-MIMO) arrays have therefore emerged as a promising approach to counteract this attenuation by concentrating energy toward desired spatial locations through beamforming. At such frequencies and aperture sizes, the propagation model departs from the conventional narrowband far-field assumption, as many terminals reside within the array’s near-field region \cite{squintluodai}. In this wideband near-field regime, beamforming must achieve true focal focusing rather than simple directional steering. A fundamental challenge then arises because subcarriers experience distinct, frequency-dependent propogation delays, resulting in beam squint \cite{squintsurvey,squintluodai}. Consequently, a beam sharply focused at the center frequency may defocus toward the band edges, leading to significant signal-to-noise ratio (SNR) degradation.

One straightforward solution is to employ a fully digital array architecture. By equipping each antenna element with a dedicated RF chain and high-speed data converter, the transmitter can perform subcarrier-level precoding to compensate for broadband distortion and mitigate multi-user interference. However, the hardware complexity grows with array size, making the per-element RF hardware prohibitively power- and cost-intensive at multi-gigahertz bandwidths and large apertures \cite{dai2022delay}.
To address this challenge, replacing phase shifters (PSs) with true-time-delay (TTD) circuits enables frequency-dependent beam steering and effectively eliminates far-field beam squint, while hybrid extensions support multi-user near-field downlink operation \cite{dai2022delay,wang2025mu}. TTD-based hybrid beamforming typically adopts either parallel or serial networks, each with inherent drawbacks. Parallel architectures provide per-path delay control but require very large absolute delays on each TTD device on the order of hundreds to thousands of picoseconds, which increases cost and hinders compact integration. Serial architectures reduce the per-device delay through accumulation but introduce strong inter-stage coupling that prevents independent control and complicates calibration \cite{dai2022delay,wang2024ttd}. Meanwhile, the element layout couples with frequency and thus shapes band-edge SNR and focusing robustness. Canonical layouts, including uniform linear arrays (ULAs)\cite{cai2025hybrid,dai2022delay,wang2025mu,wang2024ttd}, uniform circular arrays (UCAs) \cite{guo2024uca}, and uniform concentric circular arrays (UCCAs) \cite{uccawideband}, exhibit distinct frequency dependent responses. For example, ULAs generally preserve beam shape across frequency, whereas UCAs trade that stability for sharper focusing that can suppress leakage and improve spectral efficiency. In practice, element positions are fixed at fabrication, so no single static geometry is universally optimal for wideband multiuser near field downlink, because user distributions and channels vary over space and time.

This observation motivates a natural question: \emph{Can the array layout be adapted to enable wideband-friendly focusing and enhance multi-user performance without relying on TTD networks?} The emergence of next-generation reconfigurable antenna (NGRA) systems, including movable antennas (MA) system \cite{MAsurvey2} and fluid antenna (FA) system\cite{Fasurvey1}, makes this feasible. Systems can reconfigure element positions at run time, either by activating different ports or by mechanically and electromechanically shifting elements. The additional spatial degrees of freedom (DoFs) offered by MAs enable flexible beamforming and can substantially enhance multi-user communication performance. For example, controlled antenna motion can reduce self-interference and improve integrated sensing and communication (ISAC) performance \cite{maself1}, and combining MAs or FAs with multiple-access schemes can boost throughput \cite{maself2,faself1}.

These observations motivate leveraging NGRA to adapt the array layout to the channel environment. The work in \cite{mawideband} investigates a single-input single-output (SISO) low-frequency wideband OFDM system, whereas \cite{zhuyanze} considers a simple wideband multiple-input single-output (MISO) setting and optimizes MA positions. However, a systematic framework for array layout design in multi-user near-field wideband systems remains unexplored. To fill this gap, we first reveal how the array layout fundamentally influences wideband performance, and then propose an MA-aided hierarchical sub-connected hybrid beamforming (HSC-HBF) architecture along with an efficient optimization algorithm that reduces both hardware and computational complexity. Simulation results verify that the proposed approach effectively suppresses beam squint under diverse scenarios and antenna number configurations.

\textit{Notations:} $\mathbf{x}^{\mathsf T}$, $\mathbf{X}^{\mathsf H}$, $[\mathbf{x}]_{i}$, and $\|\mathbf{x}\|_{i}$ denote the transpose, the conjugate transpose, the $i$th entry of $\mathbf{x}$, and the $\ell_i$-norm, respectively.

\section{System Model}\label{system model}
We consider a wideband near-field downlink MU-MISO system illustrated in Fig.~\ref{fig:system-model}, in which a multi-antenna base station (BS) employs an MA-aided HSC-HBF architecture to serve $K$ single-antenna user equipments (UEs) indexed by $k\in\{1,\dots,K\}$.
\subsection{HSC-HBF Structure}
The array comprises \(N_{\mathrm{RF}}\) \textbf{panels}, each connected to a dedicated RF chain. Let \(N_{\mathrm{RF}}=N_{P,h}N_{P,v}\), where \(N_{P,h}\) and \(N_{P,v}\) denote the numbers of panels along the horizontal and vertical axes, respectively. With \(N\) antennas in total, each panel contains \(N_{\mathrm{sub}}=N/N_{\mathrm{RF}}\) antennas. Within a panel, the \(N_{\mathrm{sub}}\) antennas are arranged as \(N_T\) \textbf{tiles} of \(N_E\) \textbf{elements} (i.e., \(N_{\mathrm{sub}}=N_TN_E\)), forming a hierarchical element–tile–panel organization inherent to the MA-aided HSC-HBF architecture. The intra-tile geometry \(\{\boldsymbol{\delta}_i\}_{i=1}^{N_E}\) is fixed and shared across tiles, and each tile moves as a whole. We index panels by \((m,n)\) with \(m=1,\ldots,N_{P,h}\) and \(n=1,\ldots,N_{P,v}\), and denote the center of panel \((m,n)\) by \(\mathbf{c}^{m,n}=(0,\,y^{m,n},\,z^{m,n})^{\mathsf T}\). Within panel \((m,n)\), the \(t\)-th tile translates by \(\boldsymbol{\Delta}^{m,n}_t\) relative to \(\mathbf{c}^{m,n}\) for \(t=1,\ldots,N_T\), and the admissible element region is \(\mathcal{C}^{m,n}\). The absolute position of element \(i\) in tile \(t\) is
\begin{equation}
\mathbf{r}^{m,n}_{t,i}
= \mathbf{c}^{m,n}
+ \boldsymbol{\Delta}^{m,n}_t
+ \boldsymbol{\delta}_i
\in \mathcal{C}^{m,n},
\qquad i=1,\ldots,N_E.
\end{equation}
\subsection{Channel Model}
\begin{figure}[t]
    \centering
    \includegraphics[width=0.40\textwidth]{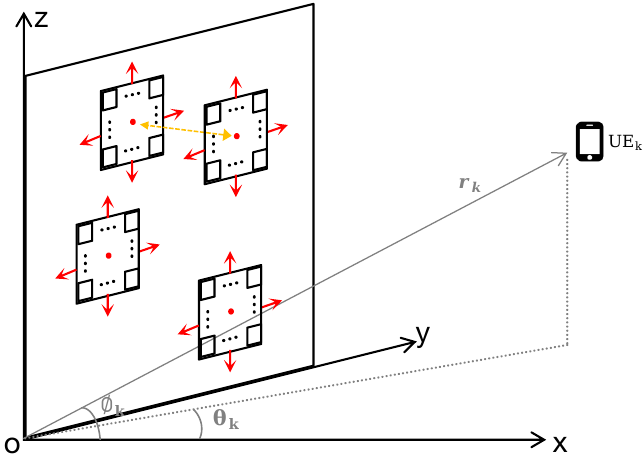}
    \caption{Illustration of coordinate frame and spatial angles.}
    \vspace{-0.5cm}
    \label{fig: distance}
\end{figure}

As illustrated in Fig.~\ref{fig: distance}, let \(r_k\) be the range from user \(k\) to the origin. Let \((\theta_k,\phi_k)\) denote the azimuth and the elevation angles, respectively. The position of user \(k\) is
\begin{equation}
\mathbf{r}_k
= \big[\, r_k\cos\phi_k\cos\theta_k,\; r_k\cos\phi_k\sin\theta_k,\; r_k\sin\phi_k \,\big]^{\mathsf T}.
\end{equation}
Let \(\mathcal{L}\!\triangleq\!\{1,\dots,L\}\) denote the subcarrier indices, \(B\) the bandwidth, and \(f_c\) the center frequency. With subcarrier spacing \(\Delta f=B/L\), the center frequency of subcarrier \(l\) is
$
f_l
= f_c+\frac{B}{2L}\,(2l-1-L),l\in\mathcal{L}.
$
In the near-field regime, the position of element \(i\) of tile \(t\) on panel \((m,n)\) is
\(\mathbf{r}^{m,n}_{t,i}=[0,\,y^{m,n}_{t,i},\,z^{m,n}_{t,i}]^{\mathsf T}\).
Under the spherical-wavefront model~\cite{squintluodai}, the exact geometric path length from user \(k\) to this element is
\begin{equation}
\label{eq:range}
\begin{aligned}
d_k(\mathbf{r}^{m,n}_{t,i})
&= \Big( r_k^{2} + \big(y^{m,n}_{t,i}\big)^{2} + \big(z^{m,n}_{t,i}\big)^{2} \\
&\qquad -\, 2r_k\!\big( y^{m,n}_{t,i}\cos\phi_k\sin\theta_k + z^{m,n}_{t,i}\sin\phi_k \big) \Big)^{1/2}.
\end{aligned}
\end{equation}
Define the spherical-wave steering vector for user \(k\) on subcarrier \(l\) as
\begin{equation}
\mathbf{b}_{l}\!\big(r_k,\theta_k,\phi_k;\mathbf{r}\big)
\triangleq
\Big[\,e^{-j \tfrac{2\pi f_{l}}{c_0}\, d_k(\mathbf{r}^{m,n}_{t,i})}\,\Big]_{(m,n,t,i)\in\mathcal{I}},
\end{equation}
where \(\mathcal{I}\) indexes all \(N=N_{\mathrm{RF}}N_TN_E\) elements, \(\mathbf{r}\) stacks \(\{\mathbf{r}^{m,n}_{t,i}\}\), and \(c_0\) is the speed of light. According to~\cite{squintluodai}, the line-of-sight (LoS) near-field channel\footnote{Equation~\eqref{eq:range} adopts the exact spherical-wave path-length model, which is valid in both near- and far-field regimes.} is
\begin{equation}
\mathbf{h}_{l,k}(\mathbf{r})
= \beta_{l,k}\,
\mathbf{b}_{l}\!\big(r_k,\theta_k,\phi_k;\mathbf{r}\big)
\in \mathbb{C}^{N},
\end{equation}
with \(\beta_{l,k}\in\mathbb{C}\) denoting the complex path loss.

\subsection{Signal Model}
The received signal at user $k$ on subcarrier $l$ is
\begin{equation}
y_{l,k}= \mathbf{h}_{l,k}^{H}(\mathbf{r}) \mathbf{A} \mathbf{D}_{l}\mathbf{s}_{l}+n_{l,k},\label{signal model}
\end{equation}
where $\mathbf{A}\in\mathbb{C}^{N\times N_{\mathrm{RF}}}$ is the frequency-independent analog precoder implemented by phase shifters,
$\mathbf{D}_{l}\!=\![\mathbf{d}_{l,1},\dots,\mathbf{d}_{l,K}]\!\in\!\mathbb{C}^{N_{\mathrm{RF}}\times K}$ is the digital baseband precoder on subcarrier $l$,
$\mathbf{s}_{l}\in\mathbb{C}^{K}$ is the symbol vector with $\mathbb{E}[\mathbf{s}_{l}\mathbf{s}_{l}^{H}]\!=\!\mathbf{I}_{K}$. The noise term $n_{l,k}\!\sim\!\mathcal{CN}(0,\sigma^{2})$ is additive white Gaussian noise (AWGN).

For a HSC-HBF architecture, each RF chain drives a disjoint subarray comprising \(N_{\mathrm{T}}\) tiles, each containing \(N_{\mathrm{E}}\) elements.
Accordingly, the analog precoder has a block-diagonal form with one column per RF chain
\begin{equation}
\mathbf{A}=\mathrm{blkdiag}\!\big(\mathbf{A}^{sub}_{1,1},\ldots,\mathbf{A}^{sub}_{N_{P,h},N_{P,v}}\big), 
\end{equation}
\begin{equation}
\mathbf{A}^{sub}_{m,n}=\mathrm{blkdiag}\!\big(\mathbf{a}^{m,n}_{1},\ldots,\mathbf{a}^{m,n}_{N_T}\big), 
\end{equation}
where $\mathbf{A}^{sub}_{m,n}\in\mathbb{C}^{N_{\mathrm{sub}}\times N_T}$ denotes the PS-based analog precoding matrix for the subarray driven by panel $(m,n)$, and $\mathbf{a}^{m,n}_{t}\in\mathbb{C}^{N_{E}\times 1}$ is the phase-shifter vector of its $t$-th tile. Thus, the SINR for decoding user $k$ on subcarrier $l$ is
\begin{equation}
\gamma_{l,k} =
\frac{\big|\mathbf{h}_{l,k}^{H}(\mathbf{r})\,\mathbf{A}\,\mathbf{d}_{l,k}\big|^{2}}
{\displaystyle\sum_{i\neq k}\big|\mathbf{h}_{l,k}^{H}(\mathbf{r})\,\mathbf{A}\,\mathbf{d}_{l,i}\big|^{2}+\sigma^{2}}.
\end{equation}
Accounting for the OFDM cyclic prefix of length \(L_{\mathrm{CP}}\), the spectral efficiency is
\begin{equation}
R_{tot} =\sum_{k=1}^KR_k= \frac{1}{L+L_{\mathrm{CP}}}\sum_{k=1}^{K}\sum_{l=1}^{L}\log_{2}\!\big(1+\gamma_{l,k}\big).
\end{equation}

\subsection{Problem Formulation}

To verify the effectiveness of the proposed HSC-HBF architecture, we jointly optimize the analog and digital precoders together with the MA array layout to maximize the spectral efficiency. Hence, the optimization problem is formulated as

\begin{subequations}\label{eq:SE-opt-A}
\begin{align}
\max_{\ \mathbf{A},\,\,\{\boldsymbol{\Delta}^{m,n}_{t}\},\{\mathbf{D}_{l}\}} \quad &\sum_{k=1}^{K}\sum_{l=1}^{L}\log_{2}\!\big(1+\gamma_{l,k}\big)
\label{eq:SE-opt-A:obj} \\
\text{s.t.}\quad &
\|\mathbf{A}\,\mathbf{D}_{l}\|_{F}^{2} \le P_{t}, \quad \forall\, l,
\label{eq:SE-opt-A:power} \\
&
|[\mathbf{a}^{m,n}_{t}]_{e}|=1, \quad \forall\, e,\ \forall\, (m,n),\ t,
\label{eq:SE-opt-A:unitmod} \\
&
\mathbf{r}^{m,n}_{t,i}\in\mathcal{C}^{m,n}, \quad \forall\, (m,n),\ t,\ i,
\label{eq:SE-opt-A:contain} \\
&
\big\|\,\boldsymbol{\Delta}^{m,n}_{t}-\boldsymbol{\Delta}^{m,n}_{t'}\,\big\|_{2}\ \ge\ D_{\min}, \nonumber\\
&\quad\quad\quad\quad\quad\quad \forall\, (m,n),\ \forall\, t\neq t',
\label{eq:SE-opt-A:spacing}
\end{align}
\end{subequations}
where \eqref{eq:SE-opt-A:power} enforces a per-subcarrier transmit power budget. \eqref{eq:SE-opt-A:unitmod} imposes the unit-modulus property on each entry of the tile phase-shifter vectors. \eqref{eq:SE-opt-A:contain} requires every element coordinate $\mathbf{r}^{m,n}_{t,i}$ to lie inside its parent panel’s admissible region $\mathcal{C}^{m,n}$.
\eqref{eq:SE-opt-A:spacing} enforces a minimum separation $D_{\min}$ between the tile translations $\boldsymbol{\Delta}^{m,n}_{t}$ and $\boldsymbol{\Delta}^{m,n}_{t'}$ within the same panel. With appropriate design, cross-panel and intra-tile spacings are inherently satisfied.

\section{Design for Wideband Near Field Precoding}
Inspired by \cite{mmhybrid,wang2025mu}, we adopt a heuristic procedure: select \(K\) out of \(N_{\mathrm{RF}}\) RF chains and assign them one-to-one to the \(K\) users. For each selected chain, jointly configure its subarray analog precoder \(A^{sub}\) and the positions of the tiles attached to that chain to maximize the assigned user’s received power along the LoS direction. After fixing \(A^{sub}\) and the tile positions, compute the digital precoder on the resulting effective channel. In this section, we first provide a physical interpretation showing that the user-perceived array gain is layout-dependent. We then address the optimization problem in \eqref{eq:SE-opt-A}, jointly designing the array layout and beamforming.

\subsection{Layout-Dependent Array Gain: Physical Interpretation}
We set the phase-shifter vector for tile \(t\) to the conjugate of its steering vector at \(f_c\). Since \(\mathbf{b}_t\) depends on the element and tile positions, the analog precoder \(A^{sub}\) is a function of the antenna layout. For clarity, consider a specific user \(k\) and assume that the RF chain connected to panel \((m,n)\) is assigned to this user. To streamline notation, we suppress the panel indices \((m,n)\) below. Let \(\boldsymbol{\Delta}_t\) be the translation of tile \(t\). Define the spherical-wave path length to element \(i\) in tile \(t\) as
$d_{k,i}(\boldsymbol{\Delta}_t)\triangleq d_k(\mathbf{r}_{t,i})$. Thus, the average array gain can be derived as follows.
\begin{equation}
\begin{aligned}
G_k\big(\{\boldsymbol{\Delta}_{t}\}\big)
&=\frac{1}{L}\sum_{l=1}^L \Bigg|\mathbf{b}_l^H(r_k,\theta_k,\phi_k;\{\boldsymbol{\Delta}_{t}\})\mathbf{A}^{sub}\Bigg|\\
&=\frac{1}{L}\sum_{l=1}^L\Bigg|\sum_{t=1}^{N_T}\sum_{i=1}^{N_e}e^{j\frac{2\pi f_ld_{k,i}(\boldsymbol{\Delta}_t)}{c_0}}a_{t,i}\Bigg|\\
&=\frac{1}{L}\sum_{l=1}^L\Bigg|\sum_{t=1}^{N_T}\sum_{i=1}^{N_e}e^{j\frac{2\pi (f_l-f_c)}{c_0}d_{k,i}(\boldsymbol{\Delta}_t)}\Bigg|.
&\\
\end{aligned}
\end{equation}
Under the fixed-position antenna (FPA) baseline, the tiles are immobile. Even with conjugate steering at \(f_c\), residual frequency-dependent phases \(\Delta\psi_{l,i}=\tfrac{2\pi(f_l-f_c)}{c_0}\,d_{k,i}\) remain, so phase alignment
deteriorates as \(f_l\) departs from \(f_c\). By contrast, adapting the layout \(\{\boldsymbol{\Delta}_t\}\) operates through the distances \(\{d_{k,i}(\boldsymbol{\Delta}_t)\}\), reshaping these residual phases and, in turn, controlling how well the signals from different tiles and elements remain phase-aligned over the whole bandwidth.


\begin{algorithm}[t]
\caption{Tile-wise HSC-HBF.}
\label{alg:tile_sca}
\begin{algorithmic}[1]
\Require Fixed intra-tile layout; initial tile centers $\{\boldsymbol{\Delta}_s\}_{s=1}^{N_T}$; inner budget $V_{\max}$; tolerance $\varepsilon$; threshold $\epsilon_J$.
\For{$t=1$ \textbf{to} $N_T$}
    \State $\boldsymbol{\Delta}_t^{(0)} \gets \boldsymbol{\Delta}_t$.
    \For{$v=1$ \textbf{to} $V_{\max}$}
        \State Compute $J_l\big(\{\boldsymbol{\Delta}_s^{(v-1)}\}\big)$ for all $l=1,\dots,L$.
        \State $J_{\min}\gets\min_{l} J_l\big(\{\boldsymbol{\Delta}_s^{(v-1)}\}\big)$.
        \State $\mathcal{L}_{\epsilon}\gets\{\,l: J_l\big(\{\boldsymbol{\Delta}_s^{(v-1)}\}\big)-J_{\min}\le\epsilon_J\,\}$.
        \State Solve the convex subproblem \eqref{eq:sca-subproblem} restricted to $l\in\mathcal{L}_{\epsilon}$ to get $\boldsymbol{\Delta}_t^{(v)}$.
        \If{$\|\boldsymbol{\Delta}_t^{(v)}-\boldsymbol{\Delta}_t^{(v-1)}\|\le\varepsilon$} 
            \State \textbf{break}.
        \EndIf
    \EndFor
    \State Accept $\boldsymbol{\Delta}_t \gets \boldsymbol{\Delta}_t^{(v)}$.
\EndFor
\State Update digital precoders by solving \eqref{eq:SE-opt-D}.
\end{algorithmic}
\end{algorithm}

\subsection{MA Layout Optimization}
To identify an appropriate array layout for user \(k\), we consider the following
wideband focusing surrogate
\begin{align}
\max_{\,\{\boldsymbol{\Delta}_{t}\}} \quad
& \sum_{l=1}^{L}
\left|
\sum_{t=1}^{N_T}\sum_{i=1}^{N_e}
e^{
j\,\frac{2\pi (f_l-f_c)}{c_0}\, d_{k,i}(\boldsymbol{\Delta}_t)
}
\right| \label{eq: all carrier single user}\\
\text{s.t.}\quad
& (\ref{eq:SE-opt-A:contain}),\;(\ref{eq:SE-opt-A:spacing}). \nonumber
\end{align}
Problem \eqref{eq: all carrier single user} captures broadband focusing but is difficult to optimize jointly with the achievable rate due to its highly nonconvex and nonseparable modulus-sum structure. 

To obtain a tractable layout update that robustly improves the near-worst tones, we replace the strict minimax with a near-worst aggregation controlled by a preset \(\epsilon_J\).
Let $J_l(\{\boldsymbol{\Delta}_t\})\triangleq 
\big|\sum_{t=1}^{N_T}\sum_{i=1}^{N_e} 
e^{j\frac{2\pi f_l}{c_0}\, d_{k,i}(\boldsymbol{\Delta}_t)}\big|$ denotes the per-subcarrier array gain in \eqref{eq: all carrier single user}, set
$
J_{\min}=\min_{1\le l\le L} J_l\big(\{\boldsymbol{\Delta}_t\}\big),
\mathcal{L}_{\epsilon}\triangleq\{\,l:J_l\big(\{\boldsymbol{\Delta}_t\}\big)-J_{\min}\le \epsilon_J\,\}.
$
We then update tiles sequentially, i.e., when updating tile \(t\), keep \(\{\boldsymbol{\Delta}_{t'}\}_{t'\ne t}\) fixed and solve
\begin{subequations}\label{eq:SE-opt-B}
\begin{align}
\max_{\boldsymbol{\Delta}_{t},\,\eta}\quad
& \eta \\[2pt]
\text{s.t.}\quad
&J_l(\boldsymbol{\Delta}_t)
\;\ge\; \eta,\quad \forall\, l\in\mathcal{L}_{\epsilon},
\label{eq:SE-opt-B:worst-l}\\
& \text{\eqref{eq:SE-opt-A:contain}, \eqref{eq:SE-opt-A:spacing}}\nonumber. 
\end{align}
\end{subequations}
\subsubsection{Convexification of \eqref{eq:SE-opt-B:worst-l}}
Since \eqref{eq:SE-opt-B:worst-l} is nonconvex, we adopt an equivalent squared form,
which preserves the inequality because both sides are nonnegative. Define $Q_l(\boldsymbol{\Delta}_t)\triangleq |J_l(\boldsymbol{\Delta}_t)|^2$, then we have
\begin{equation}
Q_l(\boldsymbol{\Delta}_t)
 \ge\ \eta^{2}, \forall\, l\in\mathcal{L}_{\epsilon}.
\end{equation}
We then apply a second-order Taylor expansion of the left-hand side around the current iterate
to obtain a tractable quadratic surrogate, which will be used within a sequential convex approximation framework. Expand $Q_l(\boldsymbol{\Delta}_t)$ at $\boldsymbol{\Delta}_t^{(v)}$, we can obtain
\begin{equation}
\begin{aligned}
Q_l(\boldsymbol{\Delta}_t)\ge&\bar{Q}_l(\boldsymbol{\Delta}_t|\boldsymbol{\Delta}_t^{(v)})= Q_l(\boldsymbol{\Delta}_t^{(v)})
  + \nabla Q_l^{\mathsf T}\!\big(\boldsymbol{\Delta}_t-\boldsymbol{\Delta}_t^{(v)}\big) \\
&\,+ \tfrac{1}{2}\big(\boldsymbol{\Delta}_t-\boldsymbol{\Delta}_t^{(v)}\big)^{\!\mathsf T}
     \mathbf{U}\,\big(\boldsymbol{\Delta}_t-\boldsymbol{\Delta}_t^{(v)}\big),
\end{aligned}
\end{equation}
where $\nabla Q_l$ is the gradient and a semidefiite matrix $\mathbf{U}$ can be obtained by spectral decomposition and negative–semidefinite projection of the Hessian. 
\begin{equation}
\nabla^{2}Q_l = \mathbf{V}\,\boldsymbol{\Lambda}\,\mathbf{V}^{\mathsf T},
\mathbf{U} \triangleq \mathbf{V}\,\boldsymbol{\Lambda}_{-}\,\mathbf{V}^{\mathsf T},\
\boldsymbol{\Lambda}_{-}\!=\!\mathrm{diag}\big(\min\{\lambda_i,0\}\big).\label{Qbar}
\end{equation}
Thus, we can use SCA method to replace \eqref{eq:SE-opt-B:worst-l} to be a convex constraint
\begin{equation}
\bar{Q}_l(\boldsymbol{\Delta}_t|\boldsymbol{\Delta}_t^{(v)})\ge \eta^2, \forall\, l\in\mathcal{L}_{\epsilon}.
\end{equation}

\subsubsection{Convexification of (\ref{eq:SE-opt-A:spacing})}
For the spacing constraint $\|\boldsymbol{\Delta}_t-\boldsymbol{\Delta}_{t'}\|_2\ge D_{\min}$, by the first-order Taylor expansion at $\boldsymbol{\Delta}_m^{(n)}$ we can obtain a concave lower bound as follows.
\begin{equation}
\label{eq:blk-8}
\|\boldsymbol{\Delta}_{t}-\boldsymbol{\Delta}_{t'}\|_2
\ \ge\
\frac{\big(\boldsymbol{\Delta}_{t}^{(n)}-\boldsymbol{\Delta}_{t'}\big)^{\!\mathsf T}}
{\big\|\boldsymbol{\Delta}_{t}^{(n)}-\boldsymbol{\Delta}_{t'}\big\|_2}
\big(\boldsymbol{\Delta}_{t}-\boldsymbol{\Delta}_{t'}\big)\ge D_{min},
 \forall t'\neq t.
\end{equation}
Thus, we can obtain the successive convex approximation (SCA) subproblem as
\begin{subequations}
\begin{align}
\max_{\boldsymbol{\Delta}_t,\,\eta}\quad \label{eq:sca-subproblem}
& \eta \\[2pt]
\text{s.t.}\quad 
& \bar{Q}_l\!\big(\{\boldsymbol{\Delta}_t\}\,\big|\,\{\boldsymbol{\Delta}_t^{(v)}\}\big)\ \ge\ \eta^2, \forall l,\\
& \frac{\big(\boldsymbol{\Delta}_{t}^{(n)}-\boldsymbol{\Delta}_{t'}\big)^{\!\mathsf T}}
{\big\|\boldsymbol{\Delta}_{t}^{(n)}-\boldsymbol{\Delta}_{t'}\big\|_2}
\big(\boldsymbol{\Delta}_{t}-\boldsymbol{\Delta}_{t'}\big)\ge D_{\min},
\forall t'\neq t,\\
& (\ref{eq:SE-opt-A:contain}).\nonumber
\end{align}
\end{subequations}
After the transformation, the problem can be efficiently solved using standard convex optimization solvers such as CVX. A block-coordinate SCA scheme is adopted, wherein the tile centers are updated sequentially across outer iterations. In each step, one tile center is optimized with all others fixed, and the procedure repeats until convergence.

\subsection{Digital beamforming Design}
Based on the MA layout design and the analog beamformer, we can obtain the equivalent channel $\bar{\mathbf{h}}_{l,k}^{H}=\mathbf{h}_{l,k}^{H}(\mathbf{r})\mathbf{A}$. Thus, the subproblem for digtal precoding can be described as follows.
\begin{subequations}\label{eq:SE-opt-D}
\begin{align}
\max_{\mathbf{D}_{l}} \quad &\sum_{k=1}^{K}\sum_{l=1}^{L}\log_{2}\!\big(1+\frac{\big|\bar{\mathbf{h}}_{l,k}^{H}\mathbf{d}_{l,k}\big|^{2}}
{\displaystyle\sum_{i\neq k}^K\big|\bar{\mathbf{h}}_{l,k}^{H}\,\mathbf{d}_{l,i}\big|^{2}+\sigma_{l,k}^{2}}\big)\label{digitalprecoder}\\
\text{s.t.}\quad &
\eqref{eq:SE-opt-A:power}.\nonumber
\end{align}
\end{subequations}
This classical formulation is solved using the weighted minimum mean-square error (WMMSE) method \cite{wmmse}. Accordingly, the above steps can be summarized in \textbf{Algorithm \ref{alg:tile_sca}}.
\subsection{Complexity Analysis}
In the proposed block coordinate SCA framework, the digital precoder design in \eqref{digitalprecoder} entails negligible cost owing to the low dimension of $\mathbf{D}_l$. Therefore, the overall complexity is dominated by the subproblem in \eqref{eq:sca-subproblem}, whose computational complexity is approximately $\mathcal{O}((L + N_T)^{1.5})$. Considering that all $N_T$ tiles are sequentially updated over $I_{\text{outer}}$ outer iterations for $K$ users, the total complexity scales as $\mathcal{O}(K I_{\text{outer}} N_T (L + N_T)^{1.5})$. Moreover, enlarging the tile layout effectively reduces the number of tiles $N_T$, thereby achieving lower overall computational complexity.

\section{Numerical Results}

To mitigate mutual coupling \cite{MAsurvey2}, the inter-element spacing is kept larger than \(\lambda_c/2\), where \(\lambda_c \triangleq c_0/f_c\). The array adopts an \(s\times s\) square-tile layout with \(N_E=s^2\) elements per tile. To ensure elements from different tiles remain at least \(\lambda_c/2\) apart, the tile center spacing is set to $d_{\mathrm{tile}}=\frac{\sqrt{2}(s-1)+1}{2}\lambda_c$. Unless stated, the allowable motion is confined to a square panel of side \(A=2\sqrt{N_TN_E}\lambda_c\). The path gains $\beta_{l,k}$ are generated from the complex Gaussian distribution \(\mathcal{CN}(0,1)\), and the signal-to-noise ratio is defined as \(\mathrm{SNR}=P_t/\sigma^2\) \cite{squintluodai}. Unless otherwise specified, simulations use \(L=256\) subcarriers at \(f_c=100~\mathrm{GHz}\) with a total bandwidth of \(B=20~\mathrm{GHz}\) and a transmit SNR of \(10~\mathrm{dB}\). The stopping thresholds for the iterative updates are set as \(\epsilon=0.001\) and \(\epsilon_J=0.01\). For the single-user case, the UE is fixed at \(r_0=5~\mathrm{m}\), \(\theta_0=\pi/3\), and \(\phi_0=\pi/6\). In the multi-user case, four UEs are simultaneously served, with positions uniformly distributed over \(r\!\in[5,15]~\mathrm{m}\) and \((\theta,\phi)\!\sim\!\mathcal{U}[-\pi/3,\pi/3]\). Two baselines are considered for comparison: (i) FPA (PS–only) and (ii) FPA+TTD as in \cite{wang2025mu}, with eight TTD branches. For the 64- and 128-antenna configurations, each tile comprises \(2\times2\) and \(4\times4\) elements, respectively.

\begin{figure}[t]
    \centering
    \includegraphics[width=0.40\textwidth]{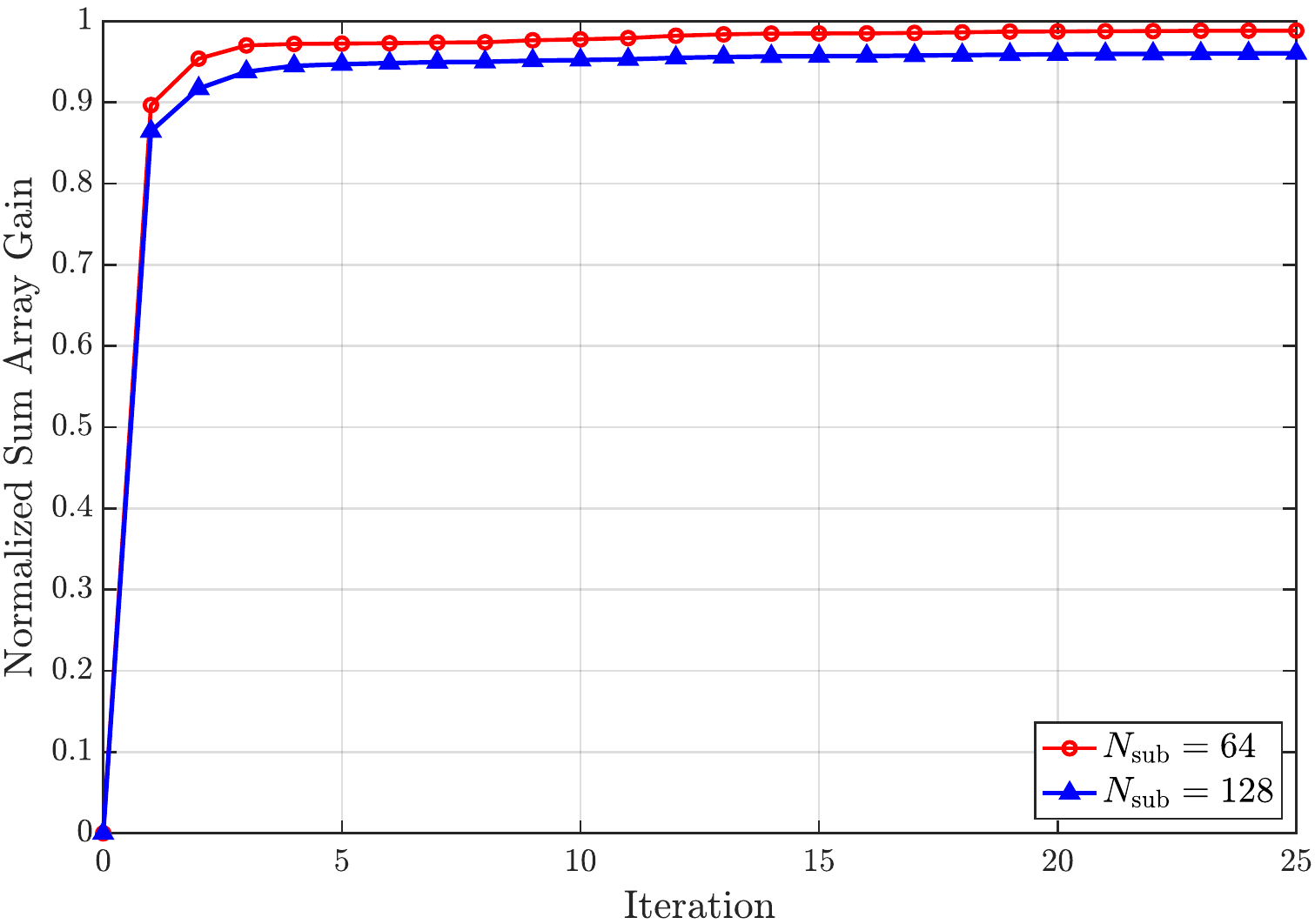}
    \caption{Normalized array gain comparison for a single user.}
    \vspace{-0.5cm}
    \label{fig:convergence}
\end{figure}

Fig.~\ref{fig:convergence} illustrates the convergence behavior of the proposed HSC-HBF algorithm in terms of the normalized sum array gain for a single-user case. It can be observed that the proposed algorithm consistently converges across different configurations, demonstrating robust stability and scalability. 

\begin{figure}[t]
  \centering
  \begin{subfigure}{0.49\linewidth}
    \centering
    \includegraphics[width=0.95\linewidth]{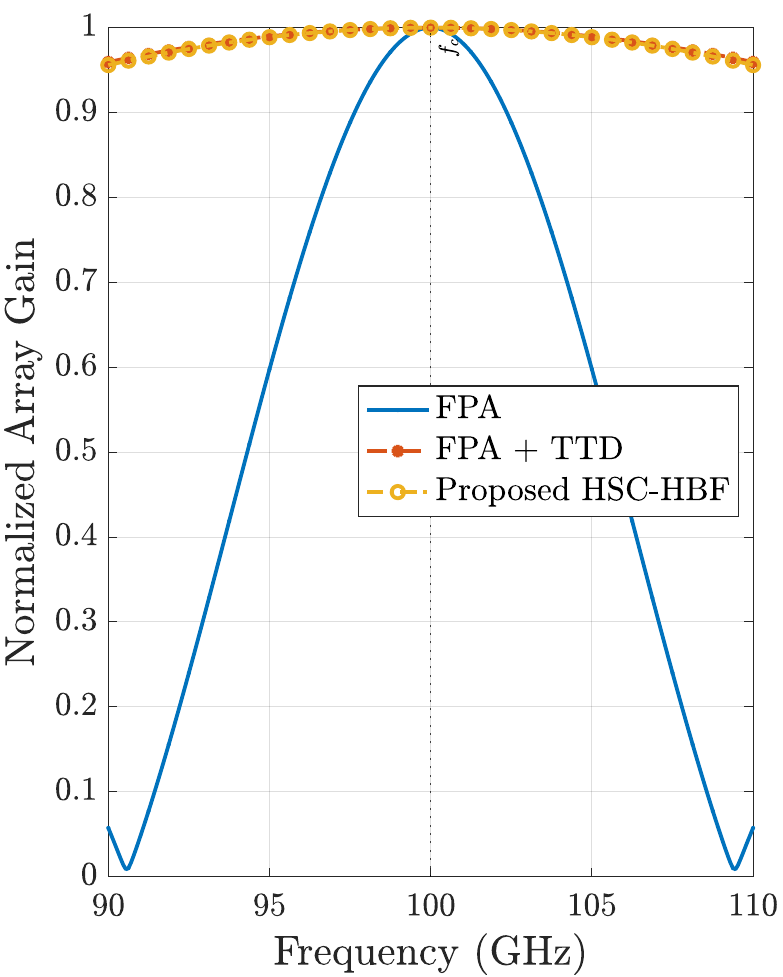}
    \caption{$N_{\text{sub}}=64$}
    \label{fig:gain_vs_f_64}
  \end{subfigure}\hfill
  \begin{subfigure}{0.49\linewidth}
    \centering
    \includegraphics[width=0.95\linewidth]{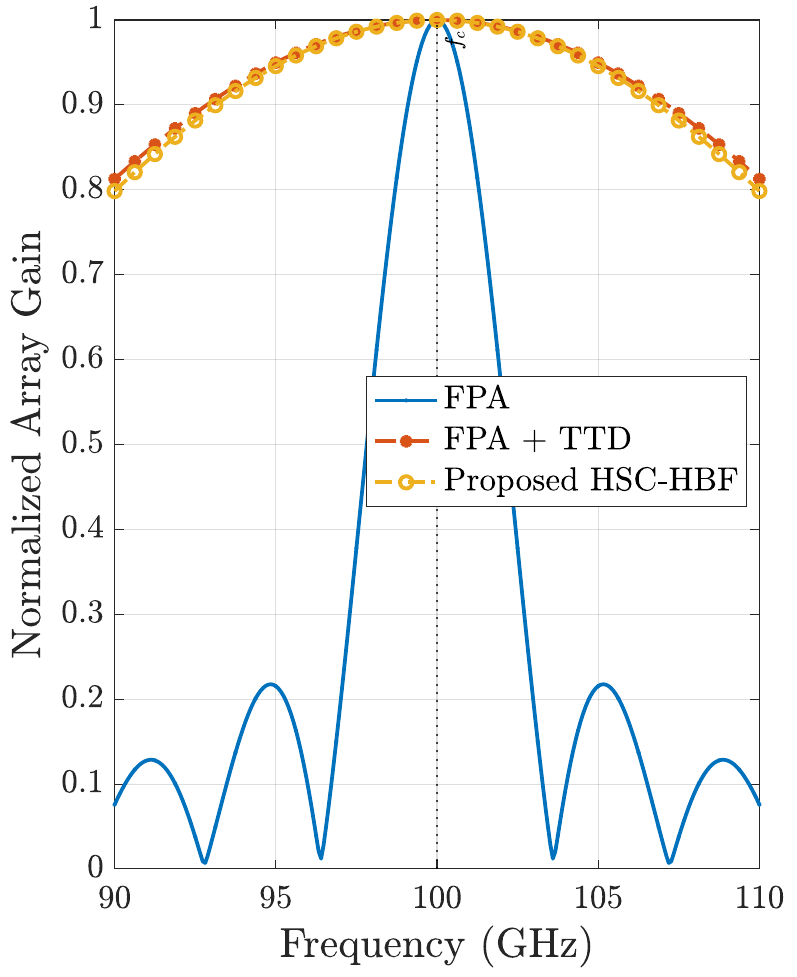}
    \caption{$N_{\text{sub}}=128$}
    \label{fig:gain_vs_f_128}
  \end{subfigure}
  \caption{Normalized array gain across different subcarrier frequencies.}
  \label{fig:gain_vs_f}
\end{figure}

Fig.~\ref{fig:gain_vs_f} presents the normalized array gain across different subcarrier frequencies for a single-user case under varing subarrays \(N_{\text{sub}}\). The proposed HSC-HBF algorithm maintains stable broadband focusing across the entire frequency band. Moreover, the HSC-HBF exhibits almost identical performance to the FPA+TTD configuration, while the FPA baseline suffers from noticeable band-edge degradation. These results demonstrate that both advanced architectures effectively suppress beam squint and achieve consistent array gain across the band.

\begin{figure}[t]
    \centering
    \includegraphics[width=0.40\textwidth]{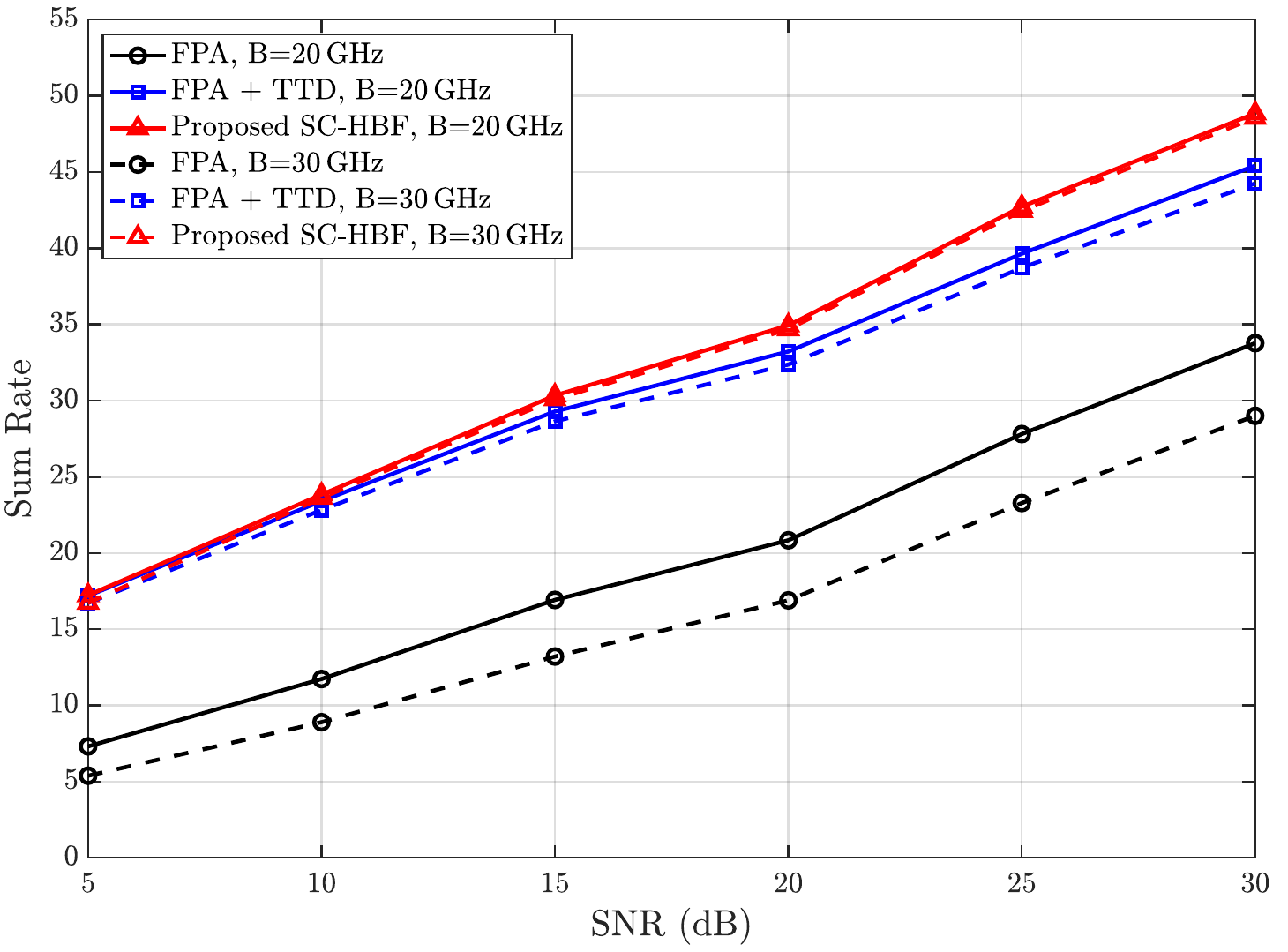}
    \caption{Sum rate across different SNR.}
    \vspace{-0.5cm}
    \label{fig:snrvsrate}
\end{figure}

Building upon the single-user results, Fig.~\ref{fig:snrvsrate} illustrates the achievable sum rate under different SNR conditions. For the single-user case, the proposed HSC-HBF architecture exhibits similar performance to the FPA+TTD configuration. However, in the multi-user scenario, the MA-aided HSC-HBF achieves a higher sum rate compared with FPA+TTD, especially at high SNRs. This improvement arises because each movable antenna’s position is adaptively designed for a specific user, resulting in weaker inter-user channel correlations, enhanced spatial separability, and reduced multi-user interference, thereby leading to better performance under high-SNR conditions. Specifically, at an SNR of 5\,dB, the proposed scheme yields 135.7\% and 210.7\% gains over FPA for 20\,GHz and 30\,GHz bandwidths, respectively, showing similar performance to FPA+TTD. When the SNR increases to 30\,dB, the HSC-HBF achieves 44.6\% and 67.4\% higher sum rates than FPA, and further improves upon FPA+TTD by 7.55\% and 9.74\%, respectively.

\begin{figure}[t]
    \centering
    \includegraphics[width=0.40\textwidth]{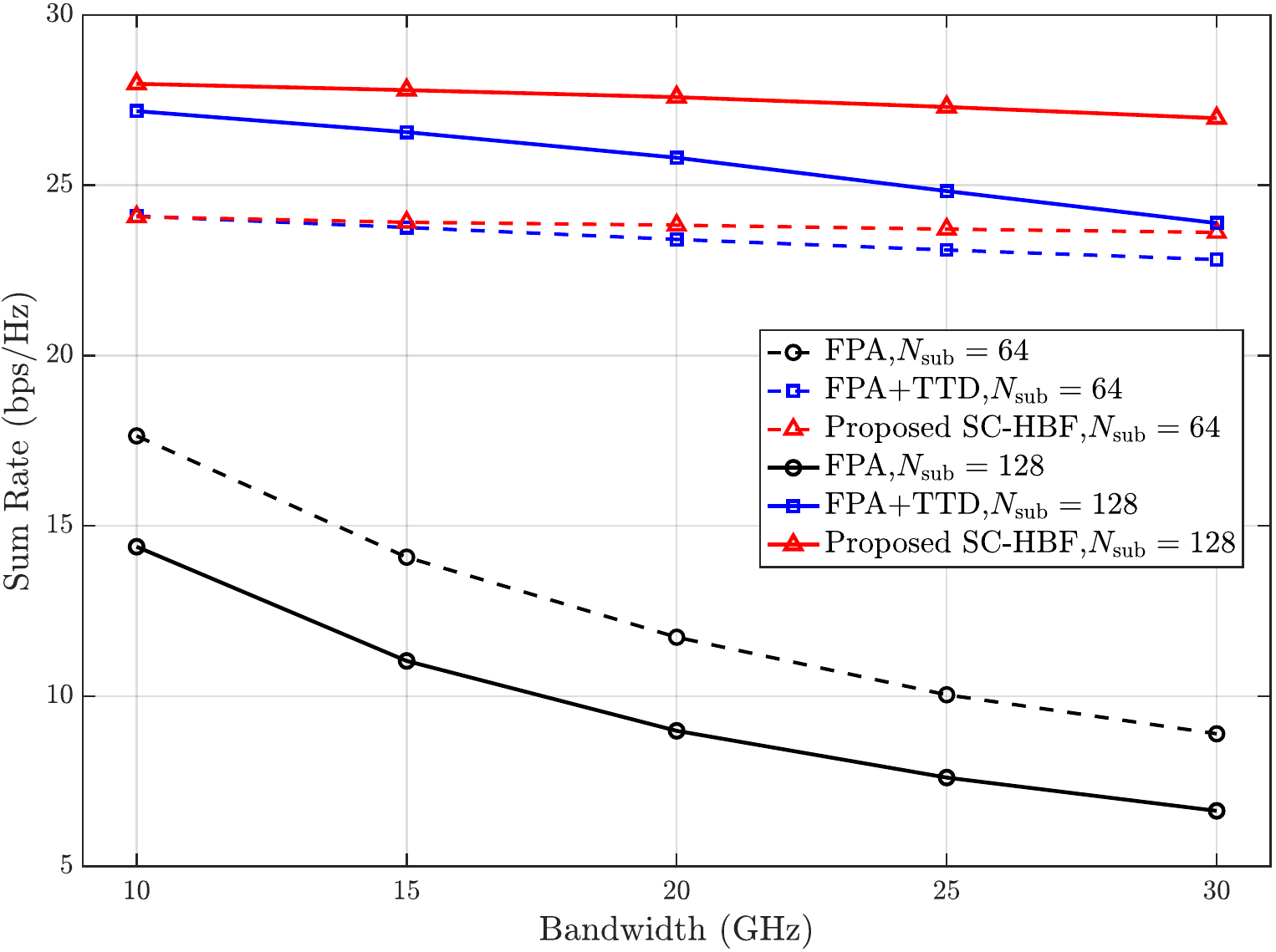}
    \caption{Sum rate across different bandwidth.}
    \vspace{-0.5cm}
    \label{fig:ratevsbandwidth}
\end{figure}

Fig.~\ref{fig:ratevsbandwidth} illustrates the sum rate performance versus bandwidth under different antenna configurations. At moderate wideband ranges, both the FPA+TTD and the proposed HSC-HBF architectures exhibit similar beam-squint mitigation performance and significantly outperform the conventional FPA array. As the bandwidth increases, all schemes experience some degradation due to frequency dispersion, while the proposed HSC-HBF maintains a more stable performance owing to its weaker inter-user channel correlations and enhanced spatial separability. At a wide bandwidth of 30\,GHz, the proposed HSC-HBF achieves 2.16\% and 5.00\% higher sum rates than FPA+TTD for \(N_{\text{sub}}=64\) and \(128\), respectively, and 78.82\% and 144.16\% improvements over FPA, demonstrating the robustness and scalability of the proposed design under wideband conditions.

\section{Conclusion}
This work proposed an MA-aided HSC-HBF architecture for wideband near-field multi-user systems. By adaptively repositioning tile centers, the design effectively mitigates beam squint without TTD networks while maintaining low hardware complexity. Simulation results verify that the proposed method achieves broadband focusing performance comparable to TTD-based schemes and delivers clear sum-rate gains over FPA. The approach offers a scalable and hardware-efficient solution for future wideband near-field communications.

\bibliographystyle{IEEEtran}
\bibliography{IEEEabrv,refs}

\end{document}